# Discussion Paper:
# Should statistics rescue mathematical modelling?

Andrea Saltelli
Centre for the Study of the Sciences and the Humanities (SVT),
University of Bergen (UIB, Bergen, Norway), and

Open Evidence Research,
Universitat Oberta de Catalunya (UOC, Barcelona, Spain)

**Abstract**

**Statistics** experiences a storm around the perceived misuse and possible abuse of its methods in the context of the so-called reproducibility crisis. The methods and styles of quantification practiced in mathematical modelling rarely make it to the headlines, though modelling practitioners writing in disciplinary journals flag a host of problems in the field. Technical, cultural and ethical dimensions are simultaneously at play in the current predicaments of both statistics and mathematical modelling.

Since mathematical modelling is not a discipline like statistics, its shortcomings risk remaining untreated longer. We suggest that the tools of statistics and its disciplinary organisation might offer a remedial contribution to mathematical modelling, standardising methodologies and disseminating good practices. Statistics could provide scientists and engineers from all disciplines with a point of anchorage for sound modelling work. This is a vast and long-term undertaking. A step in the proposed direction is offered here by focusing on the use of statistical tools for quality assurance of mathematical models. By way of illustration, techniques for uncertainty quantification, sensitivity analysis and sensitivity auditing are suggested for incorporation in statistical syllabuses and practices.

# 1 INTRODUCTION

The narrative and the order of exposition of the paper are anticipated in the present introduction. A much condensed version of the topics treated here can be found in [1].

## 1.1 DIFFERENT CRISES

Mediatic attention [2] has surrounded the emergence of a crisis in science's quality control apparatus [3]–[5]. Prima facie, the crisis concerns the non-reproducibility of scientific findings in a variety of fields such as medicine [6], [7], psychology [8], [9], economics [10], and many others [11], be it with important differences among disciplines [12]. Statistics has been especially targeted as bearing responsibility. Common accusations - to which statistics has been quick to respond [13], include poor statistical methods, poor teaching of the same, and disagreements in the profession. The angst toward statistics should not surprise; statistics is the coal face of scientific work, and hardly any scientific activity is performed without its results being couched in its language. Of course, evidence of sloppy practices is not the preserve of statistics, but affect the handling of aspects of research from cell lines to human patients [5].

When interrogated as to the causes of the crisis in statistics, statisticians are quick to point out that this is more than just a technical issue. A wrong system of incentives [14], the existence of statistical rituals [15], [16], and the delicacy of statistical work [17] have all been singled out as relevant contributors.

What is the situation in mathematical modelling? Mathematical modelling is not a discipline, and different research fields go about modelling following disciplinary habits and fashions. Modellers from various disciplines have warned that costly, dangerous, and persistent problems with mathematical modelling are connected to poor quality control and lack of appropriate standards [18], [19]. Additionally mathematical model may obfuscate the normative choice made in their construction and calibration, as is being now discussed in the controversial use of algorithms in many aspects of everyday life [20]. Without an acknowledged core, modelling lacks appropriate internal antibodies – in the form of quality standards, disciplinary associations & disciplinary journals, and recognized leaders, which would allow modellers as a constituency to issue authoritative statements such as those issued by statisticians.

An important issue seen in mathematical modelling is in the management of uncertainty. Uncertainty quantification should be at the hearth of the scientific method, and a fortiori in the use of science for policy [21]. In statistics the p-test can be misused as to overestimate the probability of having found a true effect [22]. Likewise in modelling studies certainty may be overestimated, thus producing crisp numbers precise to the third decimal digit even in situations of pervasive uncertainty or ignorance [23], including in important cases where science needs to inform policy [24]. Normative and cultural problems naturally compound the problem as for the case of statistics. It is an old refrain in mathematical modelling – first noted among hydro-geologists, that since models are often over-parametrized, they can be made to conclude everything [25].

## 1.2 STATISTICS AT THE RESCUE?

Comparing the situation of mathematical modelling with that of statistics it is natural to hypothesize that problems are likely to be more serious with the former – isn't it is after all easier to cheat with computer code than with actual samples? It is also natural to hypothesize that statistics can help mathematical modelling: modellers of all extractions are more likely to listen to statistics than to one another; next, statistical and mathematical modelling are proximate by content – both are avenues for quantification - and by pathologies. Finally, according to the nascent field of sociology of quantification, the merging of algorithms with big data blurs many existing distinctions: "*what qualities are specific to rankings, or indicators, or models, or algorithms?*" [26].

## 1.3 WHERE TO START: UNCERTAINTY QUANTIFICATION AND SENSITIVITY ANALYSIS

Mapping the route of statistics' colonization of the infinite field of mathematical modelling is beyond the remit of a single paper. But a first track can be illustrated here. If we accepts the practitioners' hunch that numerical hubris and poor accounting of the uncertainty are a key problem, then statistics can step in providing a bedrock of methodologies for uncertainty quantification and sensitivity analysis [27]. Most of the best practices in uncertainty quantification and sensitivity analysis came from - or can be framed as belonging to - statistics. Additionally, there are at the forefront of research in sensitivity analysis a large class of statistical problem in classification, variable selection and machine learning where sensitivity analysis can make an inroad which would justify and reinforce the concept that sensitivity analysis needs to be part of a statistical syllabus.

## 1.4 FROM SENSITIVITY ANALYSIS TO SENSITIVITY AUDITING

Just as per the case of statistics, no solution is possible without careful appraisal of the normative and cultural dimensions of the problem. For this reason it has been suggested – especially in relation to the use of modelling at the science-policy interface, that a technical uncertainty quantification and sensitivity analyses might be complemented by what has been named 'sensitivity auditing' [28], [29]. This new approach focuses on the modelling process, on the underlying implicit assumptions and norms, as well as on the power relationships and interests underpinning a particular model development. Though relatively young, sensitivity auditing is already mentioned in existing impact assessment guidelines [30] and in the recent report by the European Academies' association of science for policy [31]. Sensitivity auditing borrows some of its concepts from postnormal science (PNS) [32], whose contribution to a reflexive and defensible use of mathematical modelling [33] is known to practitioners [18][34]. The use of postnormal science in and for statistics would deserve a separate essay, perhaps as a continuation of the piece from [35] which notes how in adversarial settings, where the evidence is held in balance, "there is plenty of scope […] for a postnormal statistical scientist", suggesting that statisticians should be taught PNS as an essential ingredient of their training.

Given the proximity already mentioned among different kinds of quantifications, the present discussion of norms in models feeds into past [36] and present-day [37] debates on the ethical and practical implications of using metrics and algorithms [38].

## 2 STATISTICS AND ITS CRISIS

The last decade has brought to the attention of the public a crisis in science's governance and quality-control system. Here models rarely appear to forefront. The media talk about 'wrong science [2]' and 'trouble at the lab' [39], while some academicians declare most published research findings false [3]. The crisis is multifaceted, and manifests itself as the non-reproducibility of scientific findings in many fields [40]–[42] and with a surge in retractions of scientific papers [43]. A new aspect of these phenomena is that entire fields and subfields are seen as compromised - the field of empirical economics and that of nutrition being recent additions [10][44]. Other observers deny the existence of a crisis [45], while the situation is considerably complicated by the political tensions surrounding the use of science in regulation, especially in the US [46]. As a result science's methodological and ethical crisis is not unrelated from the discussion on the end of expertise and the advent of a post-truth society [47][11][48].

In statistics abuse or misuse of the p-test [13]–[15], [22], [49] has been singled out by many observers as one of the causes of the reproducibility crisis currently affecting science [2], [3], [7], [9], [50]. Misuse of p-values includes p-hacking (fishing for favourable p-values [8]) and HARKing (formulating the research Hypothesis After the Results are Known [51]). Here the desire to achieve a sought for - or simply publishable - result may lead to more or less disingenuous fiddling with the data points, the modelling assumptions, the statistical analysis, or the research hypotheses themselves [52]. It has also been noted that the issue with p-values is just the tip of the iceberg, and that all steps of statistical data analysis may suffer from poor quality statistical work [53].

Naturally, there are considerable differences among disciplines in terms of cultures and records of quality control [54], with Physics often invoked as a model to be adopted also in relation to its master of statistical techniques [55], [56].

The community of statistics exhibits a considerable latitude of opinions on the crisis – the ASA statement [13] was complemented by as many as twenty-one commentaries by top statisticians of different orientation. Statisticians also disagree as to the balance of blame – is it the publish or perish zeitgeist or is it a technical problems to be overcome with better methods [14], [57], [58]. Some point to a deeply rooted cultural problem, linked to the existence of statistical rituals and consolidated through the history of the profession [16]. Reflexive thinkers draw attention to the delicacy and complexity of statistical work, whereby different outcome can be produced even without statisticians having embarked in 'fishing expedition' to secure a desired outcome [17], [59].

The disagreements among statisticians is perhaps more apparent than real. There appear to be overall a prevailing sentiment that reliance on null hypothesis statistical testing (NHST) should be tempered by other methods as well as by a more holistic and integrative view of what constitutes evidence in each and every individual analysis [60]. A recent twist of this debate is about abolishing the concept of significance altogether [61], which has involved an unusual (in science) recourse to the instrument of a petition [62].

The problems remain severe as the crisis of science's quality control has deep roots [11]. A science which cuts corner is unfortunately ecologically fit [63] to the existing governance arrangement of scientific production [64], meaning by this that bad science tends to reproduce itself better than the virtuous sort. Additionally it appear resilient to attempts to reform [65], [66]. Still statistics appears to have taken good note of the problem and be committed to identifying solutions.

# 3 MALPRACTICES IN MATHEMATICAL MODELLING?

We have been told that in the field of clinical medical research the percentage of non-reproducible studies could reach 85% [67]. Which is that of mathematical modelling? How can the case be made that modelling needs attention, and that its predicaments are possibly more serious than those of statistics? An attempt is made here by reflecting on the vulnerability of modelling work, the existing of rituals, the opinion of practitioners, and examples of problematic use of mathematical modelling.

## 3.1 VULNERABILITY OF MODELLING

Historian Naomi Oreskes [68] has observed that model-based predictions tend to be treated as logical inferences as those met in the classic hypothetic-deductive model of science's operation. According to this model, which we brutally simplify for the purpose of the present work, laws are tested by making predictions which are then tested by experiments. If the prediction is not confirmed, then the law is 'falsified' – in the sense of proven false. For Oreskes a major problem in our use of mathematical models is in assimilating them to physical laws, and hence treating their prediction with the same dignity. This is wrong, according to Oreskes, because in order to be of value in theory testing, the model-based predictions "must be capable of refuting the theory that generated them". Since models are "complex amalgam of theoretical and phenomenological laws (and the governing equations and algorithms that represent them), empirical input parameters, and a model conceptualization" then when a model prediction fails what part of all this construct was falsified? The underlying theory? The calibration data? The system's formalization or choice of boundaries? The algorithms used in the model?

The relation between models and data is often more symbiotic than adversarial. This has been noted e.g. in climate studies, where the relation between model and data has been defined 'incestuous', exactly to make the point that in modelling studies using data to prove a model wrong is not straightforward [69].

According to Robert Rosen [70], a theoretical biologist known for his contributions to complex system biology and the definition of 'anticipatory systems' [71][72], modelling is not a science but rather a craft. For Rosen it is impossible to link 'model' and 'system modelled' within an unambiguous causal framework. In other words, given the same system, multiple representations of the system are possible, even in the context of a closed set of specifications [73], depending on the expectations and styles of the modellers.

## 3.2 RITUALS

An important analogy between statistical and mathematical modelling is in the 'ritual' use of methods. Existence of rituals in statistics has been discussed extensively by Gigerenzer [15], [16]; we similarly call cargo-cult statistics the use of statistical terms and procedures as incantations, with scant understanding of the assumptions or relevance of the calculations [58]. As per rituals in modelling the best anecdote is perhaps that offered by Kenneth Arrow. During the Second World War he was a weather officer in the US Army Air Corps working on the production of month-ahead weather forecasts, and this is how he tells the story [74]:

> *The statisticians among us subjected these forecasts to verification and they differed in no way from chance. The forecasters themselves were convinced and requested that the forecasts be discontinued. The reply read approximately like this: "The commanding general is well aware that the forecasts are no good. However, he needs them for planning purposes".*

Social scientist Niklas Luhmann, discussed in [75], uses the terms 'deparadoxification' to indicate the use of scientific knowledge to give a pretention of objectivity, to show that policy decisions are based on a publicly verifiable process, rather than on expert's whim. Along these lines it has been argued that models universally known to be wrong continue to play a role in economic policy decisions [76], which relates to our discussion of algorithms: authorities may willingly or inadvertently recur to algorithms to implements rules which - if applied by human subjects - would be either legislated or made through administrative rulemaking. Big data prediction models can be built instead and used without policy decisions ever having been the object of a political or at least traceable administrative procedure, as noted by [20].

## 3.3 PRACTITIONERS' TAKE

Its state of fragmentation has led to mathematical modelling being a field which lacks universal standards. A book on modelling read by all disciplines would go some way toward the enhancing the communication of good practices among fields. Unfortunately, such a book does not exist. A good work is that of Foster Morrison [77], though the book is silent about verification and validation procedures. These are instead well treated in [78], and in [79]. Rosen's book [70] contains deep insights from system biology - including on causality, and on 'modelling as a craft'. Still the problem is that none of the these works – independently of their merit – can command the necessary epistemic authority in the absence of a disciplinary house for modelling.

Padilla and co-workers, having run a survey involving 283 responding modellers, note that the heterogeneous nature of the modelling and simulation community prevents the emergence of consolidated paradigms, and that as a result verification and verification procedures are a rather trial and error business [19]. These authors advocate a "more structured, generalized and standardized approach to verification". Jakeman and co-workers – after reviewing several existing checklist and approaches for model quality – note that poorly informed non-modeller risk remaining "unaware of limitations, uncertainties, omissions and subjective choices in models", with consequent over-reliance in the quality of model-based inference. On their side model developers may oversimplify or overelaborate with the result of obfuscating model use. These authors [18] propose a 10 points checklist for model development. Noticeably, the list starts with the steps of definition of the model purpose and context, of conceptualization of the task, of choice of the family on models & uncertainty specification, before moving to the more usual nitty gritty issues of estimation, calibration and verification. The authors make use of model-relevant elements [33] from the tradition of postnormal science [32], including NUSAP, a system of pedigree for quantitative information [21], [80]. The list also insists on the need for a participatory approach to the construction of model, in the same epistemological tradition. In the same direction [34] take good notice of the existence of different cultures in model validation, one empirical and data driven, and another qualitative and participatory, which the authors associate to 'positivistic' versus 'relativistic' positions. These authors register the prevalence of the former view while acknowledging the usefulness of the latter especially at the stage of identifying scenarios. A discussion of this work is in [1].

### 3.4 THE CONJECTURE OF O'NEILL

A useful illustration of the analogies between statistical and mathematical modelling is offered by the conjecture of O'Neill ([81], p. 70), see Figure 1. This conjecture posits that too simple a model may miss important feature of the system, and thus lead to systematic error, while a too complex one – burdened by an excessive number of estimated parameters, may lead to a greater imprecision due the error propagation. In data science a very similar trade-off is that between under- and over-fitting. E.g., in interpolation one may use a polynomial whose order is too low, thus underfitting the existing or training points. If the order is too high the existing points are fitted too well at the cost of doing badly when new points are added to the pool. An array of methods is available in data science to tackle this problem [82]. Other disciplines might have different names for this trade off. In system analysis Zadeh calls this the principle of incompatibility, whereby as complexity increases "precision and significance (or relevance) become almost mutually exclusive characteristics" [83].

Finding the right balance – the point of minimum error in Figure 1, is of course easier in data analysis than it is in modelling – a point further discussed later. An old discussion of this conundrum between model parsimony and modelling hubris is in [25], who notes that models' over-parametrization may be fiddled with "to produce virtually any desired behaviour, often with both plausible structure and parameter values".

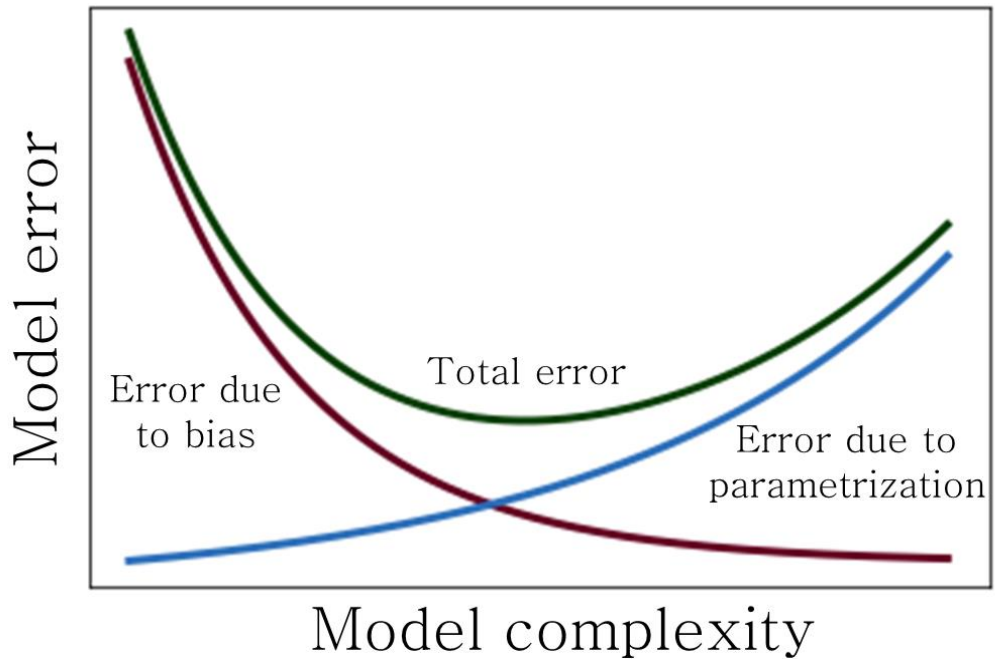

*Figure 1.* Increasing the complexity of a model reduce its systematic bias – due to the omission of relevant feature of the system being modeled. Since descriptive power comes to the price of extra estimated parameter, increasing the complexity of the model also increases the error due to the propagation of parametric uncertainty to the output.

## 3.5 REAL LIFE EXAMPLES

There are fields where the suffering in mathematical modelling has been more visibly flagged. "Useless Arithmetic" [24], argues that quantitative mathematical models used by policy makers and government administrators to form environmental policies are seriously flawed. The authors offer a host of example from AIDS prevention to the stock of fisheries, from mill tailing to costal erosion. An excerpt is offered here to give the flavour of the work (p.xiii):

> *The discredited Brunn rule model predicts how much shoreline erosion will be created by sea-level rise, and since no other model claims to do this, the Brunn rule remains in widespread use. The maximum sustainable yield is a concept that fishery models are still using as a mean to preserve fish populations despite the fact that the concept was discredited thirty-five years ago.*

### 3.5.1 Modelling of a nuclear waste disposal

One case treated in the book just mentioned is that of the nuclear waste repository at Yucca Mountain. Here a key number used in the assessment, the percolation rate of water to the unsaturated repository level - remained underestimated by 4 orders of magnitude for more than a decade [84]. Measurements made public in 1996 and later confirmed [85] revealed the presence of Chlorine-36, a bomb-pulse isotope associated to nuclear explosion at the Bikini Atoll in South Pacific 1963. Such a presence implies a travel speed of 3,000 millimetres per years, rather than the range 0.01 to 1 millimetre used in the Total System Performance Assessment model, a mathematical computer program made of 13 models in turn comprised of 286 individual modules. The modelling activity around TSPA has continued [86], tracking the fate of radionuclides thousands of years into the future. The author of the present work – likewise guilty - has computed radiological doses to population one hundred thousand years into the future (http://www.oecd-nea.org/tools/abstract/detail/nea-0860/). Today he would consider that this kind of model-based analyses reliant on the physics, the chemistry, the geology and the radio-isotopic properties of the waste are a fragile input to the safety case of a nuclear disposal, and that the surrounding institutional and social settings and power relations should be the focus of our attention. This would appear to be one of those cases where a narrowly defined scientific knowledge would give prominence to a restricted agenda of defined uncertainties. [87] in particular discuss the concept of indeterminacy – as distinguished from uncertainty - with the following example:

> *will the high quality of maintenance, inspection, operation, etc, of a risky technology be sustained in future, multiplied over replications, possibly many all over the world?*

If one contrasts the mathematical precision of TSPA with the confusion and neglect surrounding present day's nuclear waste in dump sites in the US [88] it easy to grasps the meaning of Wynne's intuition.

### 3.5.2 How much will climate cost?

The community of modellers involved in climate change is not exempt from practices which statistics would censor. Treating estimates based on an ad hoc collection of related numerical models with unknown, potentially large systematic errors as if they were a random sample from a distribution centred at the parameter is an example [58]. While modelling the future sea level and temperature appear with all their difficulties a meaningful pursuit, that of translating this in percentage of GDP point – the so called 'costing' of climate change, is by the admission of one key expert in the domain a 'terra incognita' [89]. In this terra incognita modellers venture to investigate - using large mathematical model - the increase in crime rate at county level as a result of climate change one hundred years from now [23], [90], [91]. Modellers from the same community argues for more efforts and resources in this venture [92]. Both costing climate one hundred years from now and assessing the fate of nuclear waste tens of thousands of years from now <u>using quantitative predictive models</u> resonate with what Alvin Weinberg called trans-science [93], i.e. a practice which lends itself to the language and formalism of science but where science cannot provide answers. In the specific case of attaching a cost to climate change, an uncertainty and sensitivity analysis can be shown to support this conclusion: the uncertainty of the prediction is so wide as to be useless for reaching a policy conclusion [94], however we might think that reaching one particular conclusion on this issue amounts to a moral imperative toward the planet.

### 3.5.3 Finance and economics

As per economics, conspicuous model failure has been associated with finance [95], relating to models used in the pricing of opaque financial products held partly responsible for the onset of the last recession. Philip Mirowski devotes a full chapter in his 'Never Let a Serious Crisis Go to Waste' [96] to disparage the over-reliance of economists on dynamic stochastic general equilibrium (DSGE) models used for policy simulation. This theme hit the media when both the US senate and Queen Elisabeth asked their economists for clarification about these models' apparent incapacity to anticipate the crisis [97]. Incidentally, DSGE are among the solution proposed to 'cost' climate [92]. Economist Paul Romer har recently coined the neologism 'mathiness' [98], taking issue against 'freshwater economists' (an allusion to the Chicago school in the great lakes region) for their use of mathematics as Latin, in the sense that mathematics would be used to distract from underlying ideological stances. Romer draws unflattering parallels between macroeconomic and string theory in physics for their invocation to "imaginary causal forces" and excessive deference to authority of the profession's leaders [99], and invites [100] his fellow economists to appreciate the importance of intellectual honesty by reading Richard Feynman's 'Cargo Cult' lecture, given at Caltex in 1974 [101]. The lecture famously argued that scientific integrity translates into 'utter honesty' and 'leaning over backwards' to share all details that could throw doubt on our own analysis: all the information should be given to other investigators for them to judge the value of our own contribution. More on this in the conclusions.

### 3.5.4 Algorithms of 'math destruction'?

Pathologies have been noted in the domain of algorithms – themselves models of a sort, where the use of opaquely calibrated models embedding unscrutinised rules is growing [102]. In algorithms - as well as in modelling, trade secrecy may prevent the dissemination of precisely those models which are most widely used [20]. Algorithms today decide upon an ever-increasing list of cases, such as recruiting [103], deciding on carriers - including of researchers [104], prison sentencing, paroling, custody of minors [20], with a popular book talking of "Weapons of Math Destruction" [38]. In New York, where algorithms are used by the administration for a large array of decisions, the mayor has decided to pursue legislation for "algorithmic audits" [105] and agendas are set to tackle algorithmic governance [106][1].

### 3.5.5 How many planets do we need?

The mathematical protocol developed by the Global Footprint Network (GFN) aims to assess man's impact on the planet. Simplifying, this is done by aggregating across scales and compartments different impacts due to our use of natural resource into a single number: how much of the planet's surface we use. Thus, the result of the analysis can be communicated as an overall measure of man's impact, such as the 'Earth overshot day', e.g. 'August 2 is Earth Overshoot Day 2017' (www.overshootday.org) signifies that in less than 8 months, humanity exhausts Earth's budget for the year. This number is problematic. To give an example, the EF protocol suggests that we are overshooting the capacity of the planet by about 50%, but the quantity of nitrogen used in agriculture alone would already require 2.5 planets to be fixed by natural processes – not just 1.6. The 1.6 number could become 16 or 160 or infinity, depending on what impact of man is considered – infinity corresponding to irreversible processes, e.g. man extinguishing a species by overfishing or by pesticides. Several practitioners have objected to the EF; a wealth of references can be found starting from a joint paper written by EF proponents and opponent, [107]. Also noticeable is the negative opinion about the EF in the Report by the Commission on the Measurement of Economic Performance and Social Progress led by J.

---

[1] Further references on bias in automated systems can be found at the site https://abebabirhane.wordpress.com/.

Stiglitz, JP Fitoussi and A. Sen (www.stiglitz-sen-fitoussi.fr) [108]. This has not affected the success of the Ecological Footprint, and its take-up by countries, organizations, and media. Perhaps part of the resilience of the EF is due to its complex algorithmic nature – more complex to be proven false than a non-reproducible p-value in data analysis, in line with our diagnosis that the situation with mathematical modelling is worse than in data analysis. Again, the ethical issue comes into the picture, as in the experience of the author even informed scientists may speak in favour of the EF for the desirability of the message it conveys and the mediatic effect of a crisp, single number capturing all of humanity's impact on the planet.

# 4 A MODEST BEGINNING

A step in the direction of statistics injecting some structure and standards into mathematical modelling would be by situating in the house of statistics tools for quality assurance of mathematical models. This is not a radically new idea. In the UK a group known as MUCM, for Managing Uncertainty in Complex Models, has been active in 2006-2012 on issues of uncertainty in simulation models addressing uncertainty quantification, uncertainty propagation, uncertainty analysis, sensitivity analysis, calibration (or tuning, history matching, etc.) and ensemble analysis. A quick glance at the project homepage[2] shows a strong statistical presence among its participants.

In this section an illustration is provided of how this approach might work for the specific case of sensitivity analysis and sensitivity auditing. This is just an example; it is not the scope of this paper to suggest that these methodologies can by themselves lead a thorough reformation of the field. Reforming modelling will involve both the adoption of methods in the practice of research and their insertion into teaching syllabuses. If nothing else, the present proposal could represent a low hanging fruit; sensitivity analysis (SA) already uses tools from statistics, such as Hoeffding decompositions [109], ANOVA, design of experiment[110], Bayesian formalism for partial variances and moment independent methods, and many others. Yet statistics does not normally use tools from SA. This needs to change as SA can make a good contribution to problems routinely with in statistics as discussed below.

In the following two sections we shall present sensitivity analysis and auditing in turn.

## 4.1 SENSITIVITY ANALYSIS

Sensitivity analysis (SA) finds use in a large class of applications, such as model selection, calibration, optimisation, quality assurance and many others [111]–[116]. Sensitivity analysis answers the question 'Which uncertain input factors are responsible for the uncertainty in the prediction?' SA is distinct from uncertainty analysis (UA), answering instead the question 'How uncertain is the prediction?' Different disciplines use these terms differently, though e.g. in economics, sensitivity analysis is often used to mean what we would call an uncertainty analysis.

---

[2] http://www.mucm.ac.uk/

Recent reviews of sensitivity analysis are [117]–[119], while useful textbooks for sensitivity analysis are [27], [120]–[124]. With time good practices have become consolidated such as the variance-based measures of sensitivity. These methods decompose the variance of the output into terms of the first order and terms of higher orders describing interactions. Another emerging good practice is that of in moment-independent methods [118], [125], [126]. Instead of looking at how factors affect the output variance, these methods look at how they affect the empirical probability distribution of the output as generated by the uncertainty analysis. As they look at the entire distribution and not just a moment (such as the variance) they are called moment-independent.

Sensitivity analysis is acknowledged as a useful practice in model development and application. Its use in regulatory setting – e.g. in impact assessment studies – is prescribed in guidelines both in Europe and the United States (European Commission[30], 2015, p. 390-393; Office for the Management and Budget[127], 2006, p. 17-18; Environmental Protection Agency[128], 2009, p.26).

Regrettably most modelling work does not include a sensitivity analysis, and that the majority of sensitivity analyses published in the literature are flawed to the point of irrelevance [129], [130].

A proper uncertainty analysis of the output of a mathematical model needs to map what the model does when the input factors are left free to vary over their range of existence. *A fortiori* this is true of a sensitivity analysis. Most UA and SA still explore the input space moving along mono-dimensional corridors, which leaves the space of variation of the input factors mostly unscathed. This approach in known as OAT, for moving One factor-At-a-Time, and is the one mostly preferred by modellers for a variety of practical reasons, including the ease of interpretation of the results [131]. Statisticians would suggest using instead the theory of design of experiment (DOE [110]), also because OAT ignores the existence of interactions (factors behaving synergistically), and thus the results based on OAT could dramatically underestimate the uncertainty. In a risk analysis this could imply overlooking the conditions for an accident, such as e.g. missing the conditions for a runaway reaction in a chemical reactor. This is why in the theory of experimental design factors are moved in groups, rather than OAT, to optimise the exploration of the space of the factors. Thus, most published SA fail the elementary requirement to properly explore the space of the input factors and their interplay, though this finding appears - from the personal experience of the author, discipline-dependent [129], [130].

One problem in sensitivity analysis which is clear to the author from the requests he receives is what method among the many available should be used. There are many measures for the importance of a factor, so that there can then be a different sensitivity analysis from which to arbitrarily pick. This gap urgently needs closing with a clear line of argumentation about what should be used when, and foremost why. The sensitivity measure adopted must be itself defensible.

One possible reason why proper uncertainty and sensitivity analysis of model-based inference may be eschewed is because, in their candour, these techniques may show that the inference itself is too uncertain to be of any practical use. Also to be considered is that an estimate of an investment's pay-off that gives a range from a large loss to a large gain is not what the problem owner may wish to hear. It has been observed by practitioners of different disciplines [21], [52], [132] that, under these circumstances, analysts may be tempted to 'adjust' the uncertainty in the input until the output range is narrower and conveniently located in 'gain' territory. The opposite could also be the case, where the owner of the analysis wishes uncertainties to be amplified, e.g. to the effect to deter regulatory interventions. Uncertainty can be used strategically or instrumentally, as noted in [28], and a sound sensitivity analysis can help to make up one's mind about the merits of a case, e.g. to determine where one model stands in relation to the graph of Figure 1.

It has been noted that doing modelling without sensitivity analysis is like practicing orthopaedics without x-rays. To a practitioner of uncertainty or sensitivity analysis, the idea of running a mathematical model 'just once' sounds just as weird. Why would one wish to do so when we all know that at least part of the information feeding into the model is uncertain?

As uncertainty is omnipresent in modelling, putting uncertainty at the core of model quality has a potentially unifying potential. Present computing and software capacity allow for most models to be executed repeatedly - even and especially at the model development stage. Hence one of the suggestion of the present paper is that, whenever feasible, the modelling work takes place within an ideal Monte Carlo driver for uncertainty and sensitivity analysis – e.g. the model is never run just once; instead it is systematically fed a sample of input values and returns a sample of model output for statistical treatment. If the model runs within a Monte Carlo driver then all sources of uncertainty - framing uncertainties, parametric uncertainties, and so on, can be activated simultaneously, allowing rapid and useful inferences as e.g. to what contribution each set of uncertainties does to the uncertainty in the inference. In other words: since models offers statements that are conditional on their input, this conditionality should be made explicit every time the model is used, including and foremost at the stage of model construction.

## 4.2 WHAT CAN SENSITIVITY ANALYSIS CONTRIBUTE TO STATISTICS?

One particular measure used in sensitivity analysis – known as Total Sensitivity Indices $T_i$ - can have useful applications in the field of statistics. A contribution to variable selection in regression is a recent example [133].

While statisticians are familiar with Pearson correlation ratio (which is called Sobol' first order sensitivity coefficient in sensitivity analysis), they are unfamiliar with $T_i$. These indices capture the influence of all effects (in an ANOVA sense) involving a given factor [134]. To make an example, in a system with three independent factors – a simple deterministic model of the form $Y = f(X_1, X_2, X_3)$ - the total variance can be normalized to unit and decomposed as

$$S_1 + S_2 + S_3 + S_{12} + S_{13} + S_{23} + S_{123} = 1$$

Where $S_1$ is the first order index or Pearson correlation ratio:

$$S_i = \frac{V_{X_{\sim i}}\left(E_{X_i}(Y|X_i)\right)}{V(Y)}$$

The total sensitivity index for factor $X_1$ is simply $T_1 = S_1 + S_{12} + S_{13} + S_{123}$. This index can be computed as

$$T_i = \frac{E_{X_i}\left(V_{X_{\sim i}}(Y|X_i)\right)}{V(Y)}$$

$T_i$ can be used - e.g. in variable selection in regression – as a trigger to either choose or not chose a particular regressor for the case where the output of interest $Y$ is a measure of fit of a regression model. Thus $T_i$ captures the overall effectiveness of that particular regressor, inclusive of its interactions with the choice/not choice of other regressors.

The extension of this approach to problems commonly dealt with in statistics appears promising. E.g. in problems of classification. A new type of classification score can be developed using $T_i$ to drive the selection of the variables. Likewise, when studying features importance with forests of tree; instead or randomizing individual columns (this is called permutation feature importance and corresponds to exploring one variable at a time – and hence would ignore its interactions with other variables) one can use $T_i$ as a trigger to select columns. The effectiveness of this approach is in principle superior, as we now consider the importance of more columns simultaneously – i.e. with their interactions. This can open the door to all a series of applications in model selection and validation, including to machine learning, see Table 1. This is just an example of a possible applications of a good practice in sensitivity analysis to class of problems dealt with in statistics.

Sensitivity Analysis techniques have been already used to improve the quality of composite indicators, themselves models of sort. While one application is a standard application of SA to the building of an indicator [135], another demonstrates clear inconsistencies in the way variable weights are customarily used by indicators developers, and offers corrective measures [136].

*Table 1. The dark line separating computable from unknown model needs to be broken by showing how modern SA methods can help in a large class of statistical problems*

| input | Model | output | Concerned discipline | Role of sensitivity analysis |
|---|---|---|---|---|

| The $x_i$ can be sampled from given distribution for factor $X_i$ | Perfectly known | $y$ is in known deterministically | Applied mathematics | Demonstrated |
|---|---|---|---|---|
| | Computable | $\tilde{y}$ may or may not include a stocastic component | Engineering, Computer science | |
| The $\tilde{x}_i$ are data points, observations | Unknown | $\tilde{y}$ are data points, observations | Statistics | Not-applied |

## 4.3 SENSITIVITY AUDITING

Milton Friedman has famously noted that the most significant theories - those most precious for modelling - are based on the most unrealistic of assumptions [137], [138]. This leaves open the questions of deciding what counts as a significant theory, e.g. significant to whom? Joseph Stiglitz has more soberly proposed that the beauty of models is in their acting as blinders, which by leaving a number of things out allow us to see clearly what happens with those elements which are left in [139]. Even here there must be criteria, both technical and ethical, to decide that the 'leaving out' has been done on transparently formulated normative premises and that the model is still both relevant and reasonable.

The author has heard often modellers claim the 'neutrality' of their computer codes. "Computers are impervious to the lure of power", writes a clever statisticians using computers to fight the practice of gerrymandering [140]. The confusion is here between the worth of one's cause and the neutrality of one's tool. The opinion of this article is that the technique is never neutral, and hence the use of evidence – including quantitative evidence – must be conceived being prepared for an adversarial setting, unless one's work is of very solitary academic confinement. In adversarial settings neutrality should gently morph into the relative dependence or independence from arbitrary or implausible assumptions – e.g. in the gerrymandering example, to prove to the courts that the district's boundary configuration is topologically absurd and hence must be the result of partisan interests.

Thus, an important element to improve the quality of mathematical modelling should be the extension of the technical dimension of uncertainty to the epistemic and normative dimensions. For this we suggest sensitivity auditing [28], [29]. This approach is very recent and few applications are available [141], [142], though its inclusion into the European Commission's guidelines for impact assessment [30] and in the report from the European Academies [31] appear promising. Sensitivity auditing is an extension of sensitivity analysis to cover the entire modelling process, inclusive of motivation, power relations, hidden assumptions and normative frameworks. It is based on a seven-point checklist:

Rule 1: 'Check against rhetorical use of mathematical modelling'; this rule tests if the model elucidates an issue or rather obfuscates it under a veil of math and computing power;

Rule 2: 'Adopt an "assumption hunting" attitude'; the issues here is: what was 'assumed out'? What are the tacit, pre-analytical, possibly normative assumptions underlying the analysis?

Rule 3: 'Detect pseudo-science'; the question here is to detect if the magnitude of model input uncertainties has been instrumentally downplayed (e.g. to obtain a positive inference such as e.g. "the policy will yield a benefit") or inflated (e.g. to deter action: "the impact of the policy is unclear").

Rule 4: 'Find sensitive assumptions before these find you'; this is a plea to anticipate criticism and a reminder that before publishing one's results a sensitivity analysis should be run and made available.

Rule 5: 'Aim for transparency'; stakeholders should be able to make sense of, and possibly replicate, the results of the analysis;

Rule 6: 'Do the right sums'; the analysis should not solve the wrong problem - doing the right sums is more important than doing the sums right. Here the focus is the identity and the legitimacy of the storyteller, and whether other relevant stories could or should be given.

Rule 7: 'Focus the analysis on the key question answered by the model, exploring holistically the entire space of the assumptions', see our previous discussion of sensitivity analysis.

Not all rules apply to all models. When a scientific analysis is destined to inform a policy process, the rules become a sensible guide to be developed and implemented with the collaboration of the interested stakeholder and an extended peer-community [18], [143].

Rule six on checking the narratives can be extended to a quantitative analysis of the existing frames, using a modicum of quantification to check whether some of the frames can be discarded [144]. Examples of this approach are to the Programme for International Student Assessment (PISA) by the OECD, [141], to nutrition [142], and to the Ecological Footprint [145]. Issues of trust and legitimacy in quantification as discussed e.g. in [146] are also brought to bear in this class of approaches.

In relation to the call for transparency one can note that now several journals demand to 'see the data' before accepting a paper [147] and the open-data scheme is mandatory for all the research projects financed under the H2020 framework. Could the same arrangement be achieved in mathematical modelling? Requirements to make the model available are not unheard of, see e.g. the US regulatory recommendation [127] and the EU guidelines for impact assessment [30]. 'Seeing the model' does not only mean making the model available. Journals, government agencies or regulatory body could consider asking for any modelling work a proof of uncertainty and sensitivity analysis, and - when relevant - a discussion of the model's purpose, funding, validation, assumptions, process/variable included or excluded, data used in its calibration, and so on [18], [20].

# 5 FINAL CONSIDERATIONS

Most modellers I spoke with consent that while complex calculations are now being used to support important decisions, important decisions cannot be justified on the basis of calculated results without an understanding of the associated uncertainties.

At the same time – when confronted with the considerations collected in the present text – some elements of uneasiness do emerge.

One is a deep-rooted resistance to idea of non-neutrality of mathematical modelling. There is here an important cultural and ritual element at play. That a model can be an avenue of possibly instrumental 'displacement', in the sense of moving the attention from the system to its model, as discussed in [148], is still too radical for most practitioners to contemplate. Similarly rule one of sensitivity analysis against excessive or rhetorical complexity is met with a defence of the complex models on the basis of their capacity to possibly 'surprise' the analyst. Overall it is noted that modelling is too vast an enterprise to be boxed into a single quality assurance framework. Still, I believe that material exists to give this process a good start [78], [79].

A second source of discomfort is the candour – perceived as excessive, of the methodologies advocate here. Discovering that one has to arbitrarily compress the uncertainty in the assumptions in order to obtain a useful inference [132] is something best left untried. Worse still if the sensitivity analysis were to reveal that the source of the poor (e.g. diffuse) inference is an assumption hard or impossible to nail down or to compress to a lower uncertainty. Along the same discomfort axis is that the invasive probing of sensitivity auditing may invalidate the use of models by showing the futility of, e.g., too ambitious a cost benefit analysis, or by opening the door to relativism – whereby any frame can be upheld given some sort of evidence. These doubts are countered here by rule four of sensitivity auditing – that it is better to deconstruct oneself systematically than to be deconstructed in the field. This consideration is at the core of good scientific practice, as per Robert K. Merton's principle of 'Organized Scepticism' [149], whereby all ideas must be tested and subjected to rigorous, structured community scrutiny. An exposition of this principle was given by Richard Feynman in his Cargo Cult lecture already mentioned, whereby a rigorous analyst 'bends backward' to offer to her potential inquisitors all the elements to possibly deconstruct her argument.

In econometrics Mertonian concerns are known as the 'honesty is the best policy' approach [52], and is the $10^{th}$ law of applied econometrics as formulated by Peter Kennedy: 'Thou shalt confess in the presence of sensitivity. Corollary: thou shalt anticipate criticism' [150]. This reference to sensitivity analysis reconnects us to the topic chosen as an illustration of statistics' possible rescue of mathematical modelling.

How needed is this rescue? Philosopher Jerome R. Ravetz's prophesized in 1971 that entire research fields might become diseased, (p. 179 of the second edition of [151]), and noted: *"reforming a diseased field, or arresting the incipient decline of a healthy one, is a task of great delicacy. It requires a sense of integrity, and a commitment to good work, among a significant section of the members of the field; and committed leaders with scientific ability and political skill.*" It has been argued in the present work that while statistics has been seen to possess the disciplinary arrangements and committed leaders to react to a crisis, mathematical modelling lacks one. In a recent contribution to the Royal Statistical Society magazine 'Significance' [58] a to-do list is suggested for statisticians eager to help a process of reformation of statistical practice and teaching. The present proposal adds a potential item to the list.

# 6 ACKNOWLEDGMENTS

Philip Stark from the University of Berkeley, Silvio Funtowicz from the University of Bergen (NO), and William Becker from the European Commission's Joint Research Centre offered useful comments and suggestions. The remaining errors are due to the author. This work was partially funded by a Peder Sather grant of the University of Berkeley "Mainstreaming Sensitivity Analysis and Uncertainty Auditing", awarded in June 2016.

# 7 REFERENCES

[1]  A. Saltelli, "Statistical versus mathematical modelling: a short comment," *Nat. Commun.*, vol. In press, 2019.

[2]  "How science goes wrong," *The Economist*, Oct-2013.

[3]  J. P. A. Ioannidis, "Why Most Published Research Findings Are False," *PLOS Med.*, vol. 2, no. 8, 2005.

[4]  A. Benessia *et al.*, *Science on the Verge*. Arizona State University, 2016.

[5]  R. F. Harris, *Rigor mortis : how sloppy science creates worthless cures, crushes hope, and wastes billions*. Basic Books, 2017.

[6]  C. G. Begley and L. M. Ellis, "Drug development: Raise standards for preclinical cancer research," *Nature*, vol. 483, no. 7391, pp. 531–533, Mar. 2012.

[7]  J. P. A. Ioannidis, R. Califf, R. Platt, J. Selby, D. Greenberg, and P. Neumann, "Why Most Clinical Research Is Not Useful," *PLOS Med.*, vol. 13, no. 6, p. e1002049, Jun. 2016.

[8]  D. R. Shanks *et al.*, "Romance, risk, and replication: Can consumer choices and risk-taking be primed by mating motives?," *J. Exp. Psychol. Gen.*, vol. 144, no. 6, pp. e142–e158, Dec. 2015.

[9]  Open Science Collaboration (OSC), "Estimating the reproducibility of psychological science," *Science (80-. ).*, vol. 349, no. 6251, pp. aac4716–aac4716, 2015.

[10] J. P. A. Ioannidis, T. D. Stanley, and H. Doucouliagos, "The Power of Bias in Economics Research," *Econ. J.*, vol. 127, pp. F236–F265, 2017.

[11] A. Saltelli and S. Funtowicz, "What is science's crisis really about?," *Futures*, vol. 91, pp. 5–11, 2017.

[12] D. Fanelli, "'Positive' Results Increase Down the Hierarchy of the Sciences," *PLoS One*, vol. 5, no. 4, p. e10068, Apr. 2010.

[13] R. L. Wasserstein and N. A. Lazar, "The ASA's Statement onp-Values: Context, Process, and Purpose," *Am. Stat.*, vol. 70, no. 2, pp. 129–133, 2016.

[14] J. Leek, B. B. McShane, A. Gelman, D. Colquhoun, M. B. Nuijten, and S. N. Goodman, "Five ways to fix statistics," *Nature*, vol. 551, pp. 557–559, 2017.

[15] G. Gigerenzer and J. N. Marewski, "Surrogate Science," *J. Manage.*, vol. 41, no. 2, pp. 421–440, Feb. 2015.

[16] G. Gigerenzer, "Statistical Rituals: The Replication Delusion and How We Got There," *Adv. Methods Pract. Psychol. Sci.*, vol. 1, no. 2, pp. 198–218, Jun. 2018.

[17] A. Gelman and E. Loken, "The garden of forking paths: Why multiple comparisons can be a problem, even when there is no 'fishing expedition' or 'p-hacking' and the research hypothesis was posited ahead of time," 2013.

[18] A. J. Jakeman, R. A. Letcher, and J. P. Norton, "Ten iterative steps in development and evaluation of environmental models," *Environ. Model. Softw.*, vol. 21, no. 5, pp. 602–614, 2006.

[19] J. J. Padilla, S. Y. Diallo, C. J. Lynch, and R. Gore, "Observations on the practice and profession of modeling and simulation: A survey approach," *Simulation*, vol. 94, no. 6, pp. 493–506, Oct. 2018.

[20] R. Brauneis and E. P. Goodman, "Algorithmic Transparency for the Smart City," *Yale J. Law Technol.*, vol. 20, pp. 103–176, 2018.

[21] S. Funtowicz and J. R. Ravetz, *Uncertainty and Quality in Science for Policy*. Dordrecht: Kluwer, 1990.

[22] D. Colquhoun, "An investigation of the false discovery rate and the misinterpretation of p-values," *R. Soc. Open Sci.*, vol. 1, p. 140216, 2014.

[23] A. Saltelli, P. B. Stark, W. Becker, and P. Stano, "Climate Models as Economic Guides - Scientific Challenge of Quixotic Quest?," *Issues Sci. Technol.*, vol. 31, no. 3, pp. 1–8, 2015.

[24] O. H. Pilkey and L. Pilkey-Jarvis, *Useless Arithmetic: Why Environmental Scientists Can't Predict the Future*. Columbia University Press, 2009.

[25] G. M. Hornberger and R. C. Spear, "An approach to the preliminary analysis of environmental systems," *J. Environ. Manage.*, vol. 12, no. 1, 1981.

[26] E. Popp Berman and D. Hirschman, "The Sociology of Quantification: Where Are We Now?," *Contemp. Sociol.*, vol. 47, no. 3, pp. 257–266, 2018.

[27] A. Saltelli et al., *Global sensitivity analysis : the primer*. John Wiley, 2008.

[28] A. Saltelli, Â. Guimaraes Pereira, P. van der Slujis, Jeroen, and S. Funtowicz, "What do I make of your latinorumc Sensitivity auditing of mathematical modelling," *Int. J. Foresight Innov. Policy*, vol. 9, no. 2/3/4, pp. 213–234, 2013.

[29] A. Saltelli and S. Funtowicz, "When All Models Are Wrong," *Issues Sci. Technol.*, vol. 30, no. 2, pp. 79–85, 2014.

[30] European Commission, "Guidelines on Impact Assessment - European Commission," 2015. .

[31] Science Advice for Policy by European Academies, "Making sense of science for policy under conditions of complexity and uncertainty," Berlin, 2019.

[32] S. Funtowicz and J. R. Ravetz, "Science for the post-normal age," *Futures*, vol. 25, no. 7, pp. 739–755, Sep. 1993.

[33] J. R. Ravetz, "Integrated Environmental Assessment Forum, developing guidelines for 'good practice', Project ULYSSES.," 1997.

[34] S. Eker, E. Rovenskaya, M. Obersteiner, and S. Langan, "Practice and perspectives in the validation of resource management models," *Nat. Commun.*, vol. 9, no. 1, p. 5359, Dec. 2018.

[35] J. V. Zidek, "Editorial: (Post-normal) statistical science," *J. R. Stat. Soc. Ser. A (Statistics Soc.*, vol. 169, no. 1, pp. 1–4, Jan. 2006.

[36] S. M. Bird, C. Sir David, V. T. Farewell, G. Harvey, H. Tim, and S. Peter C., "Performance indicators: good, bad, and ugly," *J. R. Stat. Soc. Ser. A (Statistics Soc.*, vol. 168, no. 1, pp. 1–27, Jan. 2005.

[37] J. Z. Muller, *The tyranny of metrics*. Princeton University Press, 2018.

[38] C. O'Neil, *Weapons of math destruction : how big data increases inequality and threatens democracy*. Random House Publishing Group, 2016.

[39] "Trouble at the Lab," *The Economist*, 2013.

[40] C. G. Begley, "Reproducibility: Six red flags for suspect work," *Nature*, vol. 497, no. 7450, pp. 433–434, May 2013.

[41] C. G. Begley and J. P. A. Ioannidis, "Reproducibility in Science: Improving the Standard for Basic and Preclinical Research," *Circ. Res.*, vol. 116, no. 1, pp. 116–126, Jan. 2015.

[42] M. Baker, "Over half of psychology studies fail reproducibility test," *Nature*, Aug. 2015.

[43] A. McCook, "A new record: Major publisher retracting more than 100 studies from cancer journal over fake peer reviews," *Retraction Watch*, Apr-2017.

[44] J. P. A. Ioannidis, "The Challenge of Reforming Nutritional Epidemiologic Research," *JAMA*, vol. 320, no. 10, p. 969, Sep. 2018.

[45] D. Fanelli, "Opinion: Is science really facing a reproducibility crisis, and do we need it to?," *Proc. Natl. Acad. Sci. U. S. A.*, vol. 115, no. 11, pp. 2628–2631, Mar. 2018.

[46] A. Saltelli, "Why science's crisis should not become a political battling ground," *Futures*, vol. 104, pp. 85–90, 2018.

[47] A. Flood, "'Post-truth' named word of the year by Oxford Dictionaries," *The Guardian*, 15-Nov-2016.

[48] A. Saltelli, "Save Science from itself, in Moedas, C., Vernos, I., Kuster, S., Nowotny, H., Saltelli, A., Mungiu-Pippidi, A., Wouter Vasbinder, A., Brooks, D.R., Cunningham, P., 2019. 'Views from a Continent in Flux.' Nature 569: 481–84.," *Nature*, vol. 569, p. 483, 2019.

[49] D. Singh Chawla, "Big names in statistics want to shake up much-maligned P value," *Nature*, vol. 548, no. 7665, pp. 16–17, Jul. 2017.

[50] J. P. A. Ioannidis, B. Forstman, I. Boutron, L. Yu, and J. Cook, "How to Make More Published Research True," *PLoS Med.*, vol. 11, no. 10, p. e1001747, Oct. 2014.

[51] N. L. Kerr, "HARKing: Hypothesizing After the Results are Known," *Personal. Soc. Psychol. Rev.*, vol. 2, no. 3, pp. 196–217, Aug. 1998.

[52] E. E. Leamer, "Tantalus on the Road to Asymptopia," *J. Econ. Perspect.*, vol. 24, no. 2, pp. 31–46, May 2010.

[53] J. Leek and R. D. Peng, "P values are just the tip of the iceberg," *Nature*, vol. 520, p. 612, 2015.

[54] D. Fanelli, M. Anderson, R. de Vries, M. Rohanizadegan, and E. Evangelou, "Do Pressures to Publish Increase Scientists' Bias? An Empirical Support from US States Data," *PLoS One*, vol. 5, no. 4, p. e10271, Apr. 2010.

[55] R. Horton, "Offline: What is medicine's 5 sigma?," *Lancet*, vol. 385, p. 1380, 2015.

[56] R. Nuzzo, "How scientists fool themselves – and how they can stop," *Nature*, vol. 526, no. 7572, pp. 182–185, Oct. 2015.

[57]   A. Saltelli and P. B. Stark, "Fixing statistics is more than a technical issue," *Nature*, vol. 553, no. 7688, pp. 281–281, Jan. 2018.

[58]   P. B. Stark and A. Saltelli, "Cargo-cult statistics and scientific crisis," *Significance*, vol. 15, no. 4, pp. 40–43, Jul. 2018.

[59]   J. R. Ravetz, "Integrity must underpin quality of statistics," *Nature*, vol. 553, no. 7688, pp. 281–281, Jan. 2018.

[60]   B. B. McShane and D. Gal, "Statistical Significance and the Dichotomization of Evidence," *J. Am. Stat. Assoc.*, vol. 112, no. 519, pp. 885–895, Jul. 2017.

[61]   V. Amrhein, S. Greenland, and B. McShane, "Scientists rise up against statistical significance," *Nature*, vol. 567, no. 7748, pp. 305–307, Mar. 2019.

[62]   A. Gelman, "'Retire Statistical Significance': The discussion.," *Blog: Statistical modelling, causal inference and social sciences*, 2019. [Online]. Available: https://statmodeling.stat.columbia.edu/2019/03/20/retire-statistical-significance-the-discussion/.

[63]   P. E. Smaldino and R. McElreath, "The natural selection of bad science," *R. Soc. Open Sci.*, vol. 3, no. 160384, 2016.

[64]   P. Mirowski, *Science-Mart, Privatizing American Science*. Harvard University Press, 2011.

[65]   J. A. Banobi, T. A. Branch, and R. Hilborn, "Do rebuttals affect future science?," *Ecosphere*, vol. 2, no. 3, pp. 1–11, Mar. 2011.

[66]   M. A. Edwards and S. Roy, "Academic Research in the 21st Century: Maintaining Scientific Integrity in a Climate of Perverse Incentives and Hypercompetition," *Environ. Eng. Sci.*, vol. 34, no. 1, pp. 51–61, Jan. 2017.

[67]   I. Chalmers and P. Glasziou, "Avoidable waste in the production and reporting of research evidence," *Lancet*, vol. 374, no. 9683, pp. 86–89, 2009.

[68]   N. Oreskes, "Why Predict? Historical Perspectives on Prediction in Earth Science," in *Prediction: Science, Decision Making, and the Future of Nature*, 2000, pp. 23–40.

[69]   P. N. Edwards, "Global climate science, uncertainty and politics: Data-laden models, model-filtered data," *Sci. Cult. (Lond).*, vol. 8, no. 4, pp. 437–472, 1999.

[70]   R. Rosen, *Life Itself: A Comprehensive Inquiry Into the Nature, Origin, and Fabrication of Life*. Columbia University Press, 1991.

[71]   T. Fuller, "Anxious relationships: The unmarked futures for post-normal scenarios in anticipatory systems," *Technol. Forecast. Soc. Change*, vol. 124, pp. 41–50, Dec. 2017.

[72]   A. H. Louie, "Robert Rosen's anticipatory systems," *Foresight*, vol. 12, no. 3, pp. 18–29, Jun. 2010.

[73]   J. C. Refsgaard, P. van der Slujis, Jeroen, J. Brown, and P. van der Keur, "A framework for dealing with uncertainty due to model structure error," *Adv. Water Resour.*, vol. 29, no. 11, pp. 1586–1597, 2006.

[74]   M. Szenberg, *Eminent economists : their life philosophies*. Cambridge University Press, 1992.

[75]   H. G. Moeller, *Luhmann explained*. Open Court Publishing Company, 2006.

[76]   F. Martin, "Unelected power: banking's biggest dilemma," *New Statemen*, 13-Jun-2018.

[77]   F. Morrison, *The art of modeling dynamic systems : forecasting for chaos, randomness, and determinism*. Dover Publications, 1991.

[78]   T. J. Santner, B. J. Williams, and W. I. Notz, *The Design and Analysis of Computer Experiments*.

Springer-Verlag, 2003.

[79] W. L. Oberkampf and C. J. Roy, *Verification and validation in scientific computing*. Cambridge University Press, 2010.

[80] P. van der Slujis, Jeroen, M. Craye, S. Funtowicz, P. Kloprogge, J. R. Ravetz, and J. Risbey, "Combining Quantitative and Qualitative Measures of Uncertainty in Model-Based Environmental Assessment: The NUSAP System," *Risk Anal.*, vol. 25, no. 2, pp. 481–492, May 2005.

[81] M. G. Turner and R. H. Gardner, "Introduction to Models," in *Landscape Ecology in Theory and Practice*, New York, NY: Springer New York, 2015, pp. 63–95.

[82] G. (Gareth M. James, D. Witten, T. Hastie, and R. Tibshirani, *An introduction to statistical learning : with applications in R*. Springer, 2017.

[83] L. Zadeh, "Outline of a New Approach to the Analysis of Complex Systems and Decision Processes," *IEEE Trans. Syst. Man. Cybern.*, vol. 3, no. 1, pp. 28–44, 1973.

[84] D. Metlay, "From Tin Roof to Torn Wet Blanket: Predicting and Observing Ground Water Movement at a Proposed Nuclear Waste Site," in *Prediction: Science, Decision Making, and the Future of Nature*, D. Sarewitz, R. A. J. Pielke, and R. Byerly, Eds. Island Press, 2000.

[85] K. Campbell, A. Wolfsberg, J. Fabryka-Martin, and D. Sweetkind, "Chlorine-36 data at Yucca Mountain: statistical tests of conceptual models for unsaturated-zone flow," *J. Contam. Hydrol.*, vol. 62–63, pp. 43–61, Apr. 2003.

[86] J. C. Helton, C. W. Hansen, and P. N. Swift, "Special Issue: Performance Assessment for the Proposed High-Level Radioactive Waste Repository at Yucca Mountain, Nevada," *Reliab. Eng. Syst. Saf.*, vol. 122, pp. 1–456, 2014.

[87] B. Wynne, "Uncertainty and environmental learning 1, 2Reconceiving science and policy in the preventive paradigm," *Glob. Environ. Chang.*, vol. 2, no. 2, pp. 111–127, Jun. 1992.

[88] F. Diaz-Maurin, "Atomic Homefront: a film about struggling to live with Manhattan Project radioactive waste," *Bulletin of the Atomic Scientists*, Jun-2018.

[89] W. D. Nordhaus, "To Slow or Not to Slow: The Economics of The Greenhouse Effect," *Econ. J.*, vol. 101, no. 407, pp. 920–937, Jul. 1991.

[90] R. Kopp, D. Rasmussen, and M. Mastrandrea, "American Climate Prospectus: Economic Risks in the United States," 2014.

[91] A. Saltelli, S. Funtowicz, M. Giampietro, D. Sarewitz, P. B. Stark, and P. van der Slujis, Jeroen, "Modelling: Climate costing is politics not science," *Nature*, vol. 532, no. 7598, pp. 177–177, Apr. 2016.

[92] N. H. (Nicholas H. Stern, *Why are we waiting? : the logic, urgency, and promise of tackling climate change*. MIT Press, 2015.

[93] A. Weinberg, "Science and trans-science," *Minerva*, vol. 10, pp. 209–222, 1972.

[94] A. Saltelli and B. d'Hombres, "Sensitivity analysis didn't help. A practitioner's critique of the Stern review," *Glob. Environ. Chang.*, vol. 20, no. 2, pp. 298–302, May 2010.

[95] P. Wilmott and D. Orrell, *The Money Formula*. Wiley & Sons, 2017.

[96] P. Mirowski, *Never Let a Serious Crisis Go to Waste: How Neoliberalism Survived the Financial Meltdown*. Verso, 2013.

[97] A. Pierce, "The Queen asks why no one saw the credit crunch coming," *The Telegraph*, 05-Nov-2008.

[98]   P. Romer, "Mathiness in the Theory of Economic Growth," *Am. Econ. Rev.*, vol. 105, no. 5, pp. 89–93, May 2015.

[99]   P. Romer, "The trouble with macroeconomics," *Am. Econ.*, vol. forthcomin, 2018.

[100]  P. Romer, "Feynman Integrity," *www.paulromer.net*, 2015. [Online]. Available: https://paulromer.net/feynman-integrity/.

[101]  R. P. (Richard P. Feynman, R. Leighton, and E. Hutchings, *Surely you're joking, Mr. Feynman!* W.W. Norton and Company, 1985.

[102]  R. Mancha and H. Ali, "The Unexamined Algorithm Is Not Worth Using," *Stanford Soc. Innov. Rev.*, no. October 11, 2017.

[103]  L. Alexander, "Is an algorithm any less racist than a human?," *The Guardian*, London, 03-Aug-2016.

[104]  Abraham C., "Turmoil rocks Canadian biomedical research community," *STAT*, Boston, 01-Aug-2016.

[105]  J. Dwyer, "Showing the Algorithms Behind New York City Services," *The New York Times*, 24-Aug-2014.

[106]  J. Danaher *et al.*, "Algorithmic governance: Developing a research agenda through the power of collective intelligence," *Big Data Soc.*, vol. 4, no. 2, pp. 1–21, 2017.

[107]  A. Galli *et al.*, "Questioning the Ecological Footprint," *Ecol. Indic.*, vol. 69, pp. 224–232, Oct. 2016.

[108]  J. E. Stiglitz, A. Sen, and J.-P. Fitoussi, "Report by the Commission on the Measurement of Economic Performance and Social Progress," *Sustain. Dev.*, vol. 12, p. 292, 2009.

[109]  G. E. B. Archer, A. Saltelli, and I. M. Sobol', "Sensitivity measures,anova-like Techniques and the use of bootstrap," *J. Stat. Comput. Simul.*, vol. 58, no. 2, pp. 99–120, May 1997.

[110]  G. E. P. Box, W. G. Hunter, Hunter, and J. Stuart, *Statistics for experimenters*. New York: Wiley & Sons, 1978.

[111]  A. Saltelli, M. Ratto, S. Tarantola, and F. Campolongo, "Update 1 of: Sensitivity Analysis for Chemical Models," *Chem. Rev.*, vol. 112, no. 5, pp. PR1–PR21, May 2012.

[112]  A. Saltelli, P. Annoni, I. Azzini, F. Campolongo, M. Ratto, and S. Tarantola, "Variance based sensitivity analysis of model output. Design and estimator for the total sensitivity index," *Comput. Phys. Commun.*, vol. 181, no. 2, pp. 259–270, Feb. 2010.

[113]  A. Saltelli, S. Tarantola, and F. Campolongo, "Sensitivity Anaysis as an Ingredient of Modeling," *Stat. Sci.*, vol. 15, no. 4, pp. 377–395, Nov. 2000.

[114]  A. Saltelli, S. Tarantola, and K. P.-S. Chan, "A Quantitative Model-Independent Method for Global Sensitivity Analysis of Model Output," *Technometrics*, vol. 41, no. 1, pp. 39–56, Feb. 1999.

[115]  A. Saltelli, "Sensitivity Analysis for Importance Assessment," *Risk Anal.*, vol. 22, no. 3, pp. 579–590, Jun. 2002.

[116]  A. Saltelli, "Making best use of model evaluations to compute sensitivity indices," *Comput. Phys. Commun.*, vol. 145, no. 2, pp. 280–297, May 2002.

[117]  J. P. Norton, "An introduction to sensitivity assessment of simulation models," *Environ. Model. Softw.*, vol. 69, no. C, pp. 166–174, Jul. 2015.

[118]  P. Wei, Z. Lu, and J. Song, "Variable importance analysis: A comprehensive review," *Reliab. Eng. Syst. Saf.*, vol. 142, pp. 399–432, 2015.

[119]  E. Borgonovo and E. Plischke, "Sensitivity analysis: A review of recent advances," *Eur. J. Oper. Res.*, vol. 248, no. 3, pp. 869–887, Feb. 2016.

[120] A. Saltelli, K. Chan, and M. Scott, *Sensitivity analysis*. Wiley, 2000.

[121] A. Saltelli, S. Tarantola, F. Campolongo, and M. Ratto, *Sensitivity Analysis in Practice*. Chichester, UK: John Wiley & Sons, Ltd, 2004.

[122] D. G. Cacuci, *Sensitivity and uncertainty analysis. Volume I, Theory*. Chapman & Hall/CRC Press, 2003.

[123] E. de. Rocquigny, N. Devictor, and S. Tarantola, *Uncertainty in industrial practice : a guide to quantitative uncertainty management*. John Wiley & Sons, 2008.

[124] E. Borgonovo, "Sensitivity Analysis. An Introduction for the Management Scientist," Springer, 2017.

[125] E. Borgonovo, "A new uncertainty importance measure," *Reliab. Eng. Syst. Saf.*, vol. 92, no. 6, pp. 771–784, Jun. 2007.

[126] E. Borgonovo, W. Castaings, and S. Tarantola, "Model emulation and moment-independent sensitivity analysis: An application to environmental modelling," vol. 34, pp. 105–115, Jun. 2012.

[127] Office of Management and Budget, "Proposed Risk Assessment Bulletin," 2006.

[128] U.S. Environmental Protection Agency (EPA), "Guidance on the Development, Evaluation, and Application of Environmental Models," 2009.

[129] F. Ferretti, A. Saltelli, and S. Tarantola, "Trends in sensitivity analysis practice in the last decade," *Sci. Total Environ.*, vol. 568, pp. 666–670, Oct. 2016.

[130] A. Saltelli *et al.*, "Why so many published sensitivity analyses are false: A systematic review of sensitivity analysis practices," *Environ. Model. Softw.*, vol. 114, pp. 29–39, Apr. 2019.

[131] A. Saltelli and P. Annoni, "How to avoid a perfunctory sensitivity analysis," *Environ. Model. Softw.*, vol. 25, no. 12, pp. 1508–1517, Dec. 2010.

[132] E. E. Leamer, "Sensitivity Analyses Would Help," *Am. Econ. Rev.*, vol. 75, no. 3, pp. 308–313, 1985.

[133] W. Becker, P. Paruolo, and A. Saltelli, "Exploring Hoover and Perez's experimental designs using global sensitivity analysis," Jan. 2014.

[134] T. Homma and A. Saltelli, "Importance measures in global sensitivity analysis of nonlinear models," *Reliab. Eng. Syst. Saf.*, vol. 52, no. 1, pp. 1–17, Apr. 1996.

[135] M. Saisana, A. Saltelli, and S. Tarantola, "Uncertainty and sensitivity analysis techniques as tools for the quality assessment of composite indicators," *J. R. Stat. Soc. Ser. A (Statistics Soc.*, vol. 168, no. 2, pp. 307–323, Mar. 2005.

[136] P. Paruolo, M. Saisana, and A. Saltelli, "Ratings and rankings: voodoo or science?," *J. R. Stat. Soc. Ser. A (Statistics Soc.*, vol. 176, no. 3, pp. 609–634, Jun. 2013.

[137] M. Friedman, *Essays in Positive Economics*. University of Chicago Press, 1953.

[138] E. S. Reinert, S. Endresen, I. Ianos, and A. Saltelli, "Epilogue: the future of economic development between utopias and dystopias," in *Handbook of Alternative Theories of Economic Development*, Edward Elgar Publishing, 2016, pp. 738–786.

[139] J. E. Stiglitz, "Rethinking Macroeconomics: What Failed, and How to Repair It," *J. Eur. Econ. Assoc.*, vol. 9, no. 4, pp. 591–645, Jan. 2011.

[140] W. K. T. Cho, "Algorithms can foster a more democratic society," *Nature*, vol. 558, no. 7711, pp. 487–487, Jun. 2018.

[141] L. Araujo, A. Saltelli, and S. V. Schnepf, "Do PISA data justify PISA-based education policy?," *Int. J. Comp. Educ. Dev.*, vol. 19, no. 1, pp. 20–34, 2017.

[142] A. Saltelli and S. Lo Piano, "Problematic quantifications: a critical appraisal of scenario making for a global 'sustainable' food production," *Food Ethics*, vol. 1, no. 2, pp. 173–179, 2017.

[143] S. N. Lane, N. Odoni, C. Landström, S. J. Whatmore, N. Ward, and S. Bradley, "Doing flood risk science differently: an experiment in radical scientific method," *Trans. Inst. Br. Geogr.*, vol. 36, no. 1, pp. 15–36, Jan. 2011.

[144] A. Saltelli and M. Giampietro, "What is wrong with evidence based policy, and how can it be improved?," *Futures*, vol. 91, pp. 62–71, Feb. 2017.

[145] M. Giampietro and A. Saltelli, "Footprints to nowhere," *Ecol. Indic.*, vol. 46, pp. 610–621, Nov. 2014.

[146] T. M. Porter, *Trust in Numbers: The Pursuit of Objectivity in Science and Public Life*. Princeton University Press, 1996.

[147] R. D. Morey *et al.*, "The Peer Reviewers' Openness Initiative: incentivizing open research practices through peer review," *R. Soc. Open Sci.*, vol. 3, p. 150547, 2016.

[148] S. Rayner, "Uncomfortable knowledge: the social construction of ignorance in science and environmental policy discourses," *Econ. Soc.*, vol. 41, no. 1, pp. 107–125, Feb. 2012.

[149] R. Merton, *The sociology of science: Theoretical and empirical investigations*. 1973.

[150] P. Kennedy, *A guide to econometrics*. Wiley-Blackwell; 6 edition, 2008.

[151] J. R. Ravetz, *Scientific knowledge and its social problems*. Oxford University Press, 1971.